\RequirePackage{lineno}

\documentclass[10pt,
  aps,
  prl,
  twocolumn,
  superscriptaddress,
  showpacs,
  altaffilletter,
nofootinbib, 
  amsmath,
  amssymb
]{revtex4-2}

\usepackage{graphicx}\usepackage{bm}\usepackage[colorlinks = true, linkcolor = blue,]{hyperref}
\usepackage{multirow} \usepackage{siunitx} \usepackage{hepunits}
\usepackage{xcolor}
\usepackage[normalem]{ulem}
\usepackage{orcidlink}

\usepackage{comment} 

\usepackage{xspace}
\usepackage{upgreek}
\usepackage[nolist]{acronym} \usepackage[nodayofweek,level]{datetime}

\usepackage{ifthen}

\begin{document}
\setlength{\abovedisplayskip}{5pt}   \setlength{\belowdisplayskip}{5pt}

\newcommand{\fix}[1]{\textcolor{red}{#1}\xspace}
\newcommand{\needref}{[\textcolor{red}{ref}]\xspace}
\newcommand{\amu}{\ensuremath{a^{}_{\mu}}\xspace}
\renewcommand{\ae}{\ensuremath{a^{}_{e}}\xspace}
\newcommand{\gm}{\ensuremath{g\!-\!2}\xspace}
\newcommand{\gmtwo}{\ensuremath{g\!-\!2}\xspace}
\newcommand{\wa}{\ensuremath{\omega_{a}}\xspace}
\newcommand{\wam}{\ensuremath{\omega_{a}^{m}}\xspace}
\renewcommand{\wp}{\ensuremath{\omega_{p}}\xspace}
\newcommand{\oa}{\ensuremath{\omega^{}_a}\xspace}
\newcommand{\ocycl}{\ensuremath{\omega^{}_c}\xspace}
\newcommand{\oS}{\ensuremath{\omega^{}_S}\xspace}
\newcommand{\op}{\ensuremath{\omega^{}_p}\xspace}
\newcommand{\opmeas}{\omega_{p}^{\pp}\xspace}
\newcommand{\opprime}{\ensuremath{\omega'^{}_p}\xspace}
\newcommand{\opprimeatTexp}{\ensuremath{\omega'^{}_p(\Tr)}\xspace}
\newcommand{\opprimeatTexpprime}{\ensuremath{\omega'^{}_p(\Trprime)}\xspace}
\newcommand{\opprimetildeatTexpprime}{\ensuremath{\tilde{\omega}'^{}_p(\Trprime)}\xspace}

\newcommand{\opprimetilde}{\ensuremath{\tilde{\omega}'^{}_p}\xspace}
\newcommand{\opprimetildeofT}{\ensuremath{\tilde{\omega}'^{}_p(T)}\xspace}
\newcommand{\opprimetildeatTexp}{\ensuremath{\tilde{\omega}'^{}_p(\Tr)}\xspace}
\newcommand{\opprimetildeatTexpnew}{\ensuremath{\tilde{\omega}'^{}_p(\Trnew)}\xspace}
\newcommand{\opprimetildebold}{\ensuremath{\bm{\tilde{\omega}'^{}_p}}\xspace}
\newcommand{\vecr}{\ensuremath{\vec{r}}\xspace}
\newcommand{\weightedfield}{\ensuremath{\langle \omega'^{}_p \times M \rangle}\xspace}
\newcommand{\weightedfieldbold}{\ensuremath{\bm{f_{\rm calib}\,{\cdot}\,\langle \omega'^{}_p \times M \rangle}}\xspace}
\newcommand{\optilde}{\ensuremath{\tilde{\omega}^{}_p}\xspace}
\newcommand{\omegap}{\op}
\newcommand{\omegapfree}{\ensuremath{\omega_p^\mathrm{free}}\xspace}
\newcommand{\fid}{FID\xspace}
\newcommand{\BNLexperiment}{\ac{BNL} E821 experiment\xspace}
\newcommand{\FNALexperiment}{\ac{FNAL} Muon \gm Experiment\xspace}
\newcommand{\RunOne}{Run-1\xspace}
\newcommand{\RunFourFiveSix}{Run-4/5/6\xspace}
\newcommand{\RunTwo}{Run-2\xspace}
\newcommand{\RunThree}{Run-3\xspace}
\newcommand{\RunTwoThree}{Run-2/3\xspace}
\newcommand{\RunOneTwoThree}{Run-1/2/3\xspace}
\newcommand{\RunOneToSix}{Run-1-6\xspace}
\newcommand{\RunThreeA}{Run-3a\xspace}
\newcommand{\RunThreeB}{Run-3b\xspace}
\newcommand{\RunFour}{Run-4\xspace}
\newcommand{\RunOneA}{Run-1a\xspace}
\newcommand{\RunOneB}{Run-1b\xspace}
\newcommand{\RunOneC}{Run-1c\xspace}
\newcommand{\RunOneD}{Run-1d\xspace}
\newcommand{\RunFive}{Run-5\xspace}
\newcommand{\RunSix}{Run-6\xspace}
\newcommand{\hethree}{\ensuremath{^3\text{He}}\xspace}
\newcommand{\water}{\ensuremath{\text{H_2O}}\xspace}
\newcommand{\pp}{\mathrm{cp}\xspace}

\newcommand{\opprimeq}{\ensuremath{\omega'^{}_{p,q}}\xspace}
\newcommand{\opprimej}{\ensuremath{\omega'^{}_{p,j}}\xspace}
\newcommand{\opprimeqj}{\ensuremath{\omega'^{}_{p,q,j}}\xspace}

\newcommand{\opprimetildeq}{\ensuremath{\tilde\omega'^{}_{p,q}}\xspace}
\newcommand{\opprimetildeqj}{\ensuremath{\tilde\omega'^{}_{p,q,j}}\xspace}
\newcommand{\Tr}{\ensuremath{T^{}_{r}}\xspace}
\newcommand{\Trprime}{\ensuremath{T'_{r}}\xspace}
\newcommand{\Trnew}{\ensuremath{T^*_{r}}\xspace}

\newcommand{\sigmamuq}{\sigma^\mu_{q}}
\newcommand{\sigmamuqj}{\sigma^\mu_{q,j}}

\newcommand{\mi}{\ensuremath{m^{}_{i}}\xspace}
\newcommand{\mone}{\ensuremath{m^{}_{1}}\xspace}
\newcommand{\mtwo}{\ensuremath{m^{}_{2}}\xspace}
\newcommand{\mthree}{\ensuremath{m^{}_{3}}\xspace}
\newcommand{\mfour}{\ensuremath{m^{}_{4}}\xspace}
\newcommand{\mfive}{\ensuremath{m^{}_{5}}\xspace}
\newcommand{\msix}{\ensuremath{m^{}_{6}}\xspace}
\newcommand{\mseven}{\ensuremath{m^{}_{7}}\xspace}
\newcommand{\meight}{\ensuremath{m^{}_{8}}\xspace}
\newcommand{\mnine}{\ensuremath{m^{}_{9}}\xspace}
\newcommand{\mten}{\ensuremath{m^{}_{10}}\xspace}
\newcommand{\meleven}{\ensuremath{m^{}_{11}}\xspace}
\newcommand{\mtwelve}{\ensuremath{m^{}_{12}}\xspace}
\newcommand{\mthirteen}{\ensuremath{m^{}_{13}}\xspace}

\newcommand{\Rmu}{\ensuremath{{\mathcal R}_\mu}\xspace}
\newcommand{\Rmuprime}{\ensuremath{{\mathcal R}'^{}_\mu}\xspace}

\newcommand{\RmuprimeOne}{\ensuremath{{\mathcal R}'^{}_{\mu}(\text{Run-1})}\xspace}
\newcommand{\RmuprimeTwoThree}{\ensuremath{{\mathcal R}'^{}_{\mu}(\text{Run-2/3})}\xspace}
\newcommand{\RmuprimeOneTwoThree}{\ensuremath{{\mathcal R}'^{}_{\mu}(\text{Run-1/2/3})}\xspace}

\newcommand\Tstrut{\rule{0pt}{2.6ex}}         \newcommand\Bstrut{\rule[-0.9ex]{0pt}{0pt}}   

\newcommand{\rmagic}{\ensuremath{r_{0}}\xspace}
\newcommand{\cmagic}{\ensuremath{c_{0}}\xspace}
\newcommand{\pmagic}{\ensuremath{p_{0}}\xspace}
\newcommand{\gammamagic}{\ensuremath{\gamma_{0}}\xspace}
\newcommand{\betamagic}{\ensuremath{\beta_{0}}\xspace}
\newcommand{\dipbfield}{\ensuremath{B_{0}}\xspace}
\newcommand{\runone}{Run-1\xspace}
\newcommand{\runtwo}{Run-2\xspace}
\newcommand{\runthree}{Run-3\xspace}
\newcommand{\runfour}{Run-4\xspace}
\newcommand\runonea{Run-1a\xspace}
\newcommand\runoneb{Run-1b\xspace}
\newcommand\runonec{Run-1c\xspace}
\newcommand\runoned{Run-1d\xspace}

\newcommand{\precession}{precession-run1}
\newcommand{\field}{field-run1}
\newcommand{\BD}{BD-run1}

\newcommand{\simon}[2]{{\color{magenta}\sout{#1}#2}}
\newcommand{\simonr}[2]{{#2}}
\newcommand{\lkg}[2]{{\color{orange}\sout{#1}#2}}
\newcommand{\lkgr}[2]{{#2}}
\newcommand{\peter}[2]{{\color{blue}\sout{#1}#2}}

 \newcommand{\RmuFNALprecisionppm}{{\SI{0.125}{ppm}}\xspace}
\newcommand{\RmuFNALprecisionppb}{{\SI{125}{ppb}}\xspace}

\newcommand{\RmuprecisionFourFiveSixppm}{{\SI{0.137}{ppm}}\xspace}
\newcommand{\RmuprecisionFourFiveSixppb}{{\SI{137}{ppb}}\xspace}

\newcommand{\Rmuworldprecisionppm}{\SI{0.122}{ppm}\xspace}
\newcommand{\Rmuworldprecisionppb}{{\SI{122}{ppb}}\xspace}

\newcommand{\amuprecisionFourFiveSixppm}{{\SI{0.139}{ppm}}\xspace}
\newcommand{\amuprecisionFourFiveSixppb}{{\SI{139}{ppb}}\xspace}

\newcommand{\amuFNALprecisionppm}{\SI{0.127}{ppm}\xspace}
\newcommand{\amuFNALprecisionppb}{\SI{127}{ppb}\xspace}

\newcommand{\amuworldprecisionppm}{\SI{0.124}{ppm}\xspace}
\newcommand{\amuworldprecisionppb}{\SI{124}{ppb}\xspace}

\newcommand{\amuRunFourFiveSix}{\ensuremath{116\,592\,0710(162) \times 10^{-12}}\xspace}

\newcommand{\amuFNAL}{
{\ensuremath{116\,592\,0705(148) \times 10^{-12}}\xspace}}

\newcommand{\amuworld}{\ensuremath{116\,592\,0715(145) \times 10^{-12}}\xspace}

\newcommand{\statchangeFNAL}{{1.87}\xspace} \newcommand{\totalprecisionchangeworld}{{1.5}\xspace}

\newcommand{\QEDref}{Aoyama:2012wk,Volkov:2019phy,Volkov:2024yzc,Aoyama:2024aly,Parker:2018vye,Morel:2020dww,Fan:2022eto}

\newcommand{\EWref}{Czarnecki:2002nt,Gnendiger:2013pva,Ludtke:2024ase,Hoferichter:2025yih}

\newcommand{\latticeHVPref}{RBC:2018dos,Giusti:2019xct,Borsanyi:2020mff,Lehner:2020crt,Wang:2022lkq,Aubin:2022hgm,Ce:2022kxy,ExtendedTwistedMass:2022jpw,RBC:2023pvn,Kuberski:2024bcj,Boccaletti:2024guq,Spiegel:2024dec,RBC:2024fic,Djukanovic:2024cmq,ExtendedTwistedMass:2024nyi,MILC:2024ryz,Bazavov:2024eou}

\newcommand{\latticeHLbLref}{Blum:2019ugy,Chao:2021tvp,Chao:2022xzg,Blum:2023vlm,Fodor:2024jyn}

\newcommand{\dataHLbLref}{Colangelo:2015ama,Masjuan:2017tvw,Colangelo:2017fiz,Hoferichter:2018kwz,Eichmann:2019tjk,Bijnens:2019ghy,Leutgeb:2019gbz,Cappiello:2019hwh,Masjuan:2020jsf,Bijnens:2020xnl,Bijnens:2021jqo,Danilkin:2021icn,Stamen:2022uqh,Leutgeb:2022lqw,Hoferichter:2023tgp,Hoferichter:2024fsj,Estrada:2024cfy,Ludtke:2024ase,Deineka:2024mzt,Eichmann:2024glq,Bijnens:2024jgh,Hoferichter:2024bae,Holz:2024diw,Cappiello:2025fyf}

\newcommand{\HVPref}{RBC:2018dos,Giusti:2019xct,Borsanyi:2020mff,Lehner:2020crt,Wang:2022lkq,Aubin:2022hgm,Ce:2022kxy,ExtendedTwistedMass:2022jpw,RBC:2023pvn,Kuberski:2024bcj,Boccaletti:2024guq,Spiegel:2024dec,RBC:2024fic,Djukanovic:2024cmq,ExtendedTwistedMass:2024nyi,MILC:2024ryz,Bazavov:2024eou,Keshavarzi:2019abf,DiLuzio:2024sps,Kurz:2014wya}

\newcommand{\HLbLref}{Colangelo:2015ama,Masjuan:2017tvw,Colangelo:2017fiz,Hoferichter:2018kwz,Eichmann:2019tjk,Bijnens:2019ghy,Leutgeb:2019gbz,Cappiello:2019hwh,Masjuan:2020jsf,Bijnens:2020xnl,Bijnens:2021jqo,Danilkin:2021icn,Stamen:2022uqh,Leutgeb:2022lqw,Hoferichter:2023tgp,Hoferichter:2024fsj,Estrada:2024cfy,Ludtke:2024ase,Deineka:2024mzt,Eichmann:2024glq,Bijnens:2024jgh,Hoferichter:2024bae,Holz:2024diw,Cappiello:2025fyf,Colangelo:2014qya,Blum:2019ugy,Chao:2021tvp,Chao:2022xzg,Blum:2023vlm,Fodor:2024jyn}

\newcommand{\SMref}{Aoyama:2012wk,Volkov:2019phy,Volkov:2024yzc,Aoyama:2024aly,Parker:2018vye,Morel:2020dww,Fan:2022eto,Czarnecki:2002nt,Gnendiger:2013pva,Ludtke:2024ase,Hoferichter:2025yih,RBC:2018dos,Giusti:2019xct,Borsanyi:2020mff,Lehner:2020crt,Wang:2022lkq,Aubin:2022hgm,Ce:2022kxy,ExtendedTwistedMass:2022jpw,RBC:2023pvn,Kuberski:2024bcj,Boccaletti:2024guq,Spiegel:2024dec,RBC:2024fic,Djukanovic:2024cmq,ExtendedTwistedMass:2024nyi,MILC:2024ryz,Bazavov:2024eou,Keshavarzi:2019abf,DiLuzio:2024sps,Kurz:2014wya,Colangelo:2015ama,Masjuan:2017tvw,Colangelo:2017fiz,Hoferichter:2018kwz,Eichmann:2019tjk,Bijnens:2019ghy,Leutgeb:2019gbz,Cappiello:2019hwh,Masjuan:2020jsf,Bijnens:2020xnl,Bijnens:2021jqo,Danilkin:2021icn,Stamen:2022uqh,Leutgeb:2022lqw,Hoferichter:2023tgp,Hoferichter:2024fsj,Estrada:2024cfy,Deineka:2024mzt,Eichmann:2024glq,Bijnens:2024jgh,Hoferichter:2024bae,Holz:2024diw,Cappiello:2025fyf,Colangelo:2014qya,Blum:2019ugy,Chao:2021tvp,Chao:2022xzg,Blum:2023vlm,Fodor:2024jyn}
 
\title{Measurement of the Positive Muon Anomalous Magnetic Moment to \amuFNALprecisionppb}

\renewcommand{\thefootnote}{\fnsymbol{footnote}}
\affiliation{Argonne National Laboratory, Lemont, Illinois, USA}
\affiliation{Boston University, Boston, Massachusetts, USA}
\affiliation{Brookhaven National Laboratory, Upton, New York, USA}
\affiliation{Budker Institute of Nuclear Physics, Novosibirsk, Russia}
\affiliation{Center for Axion and Precision Physics (CAPP) / Institute for Basic Science (IBS), Daejeon, Republic of Korea}
\affiliation{Cornell University, Ithaca, New York, USA}
\affiliation{Fermi National Accelerator Laboratory, Batavia, Illinois, USA}
\affiliation{INFN, Laboratori Nazionali di Frascati, Frascati, Italy}
\affiliation{INFN, Sezione di Napoli, Naples, Italy}
\affiliation{INFN, Sezione di Pisa, Pisa, Italy}
\affiliation{INFN, Sezione di Roma Tor Vergata, Rome, Italy}
\affiliation{INFN, Sezione di Trieste, Trieste, Italy}
\affiliation{Department of Physics and Astronomy, James Madison University, Harrisonburg, Virginia, USA}
\affiliation{Institute of Physics and Cluster of Excellence PRISMA+, Johannes Gutenberg University Mainz, Mainz, Germany}
\affiliation{Department of Physics, Korea Advanced Institute of Science and Technology (KAIST), Daejeon, Republic of Korea}
\affiliation{Lancaster University, Lancaster, United Kingdom}
\affiliation{Michigan State University, East Lansing, Michigan, USA}
\affiliation{North Central College, Naperville, Illinois, USA}
\affiliation{Northern Illinois University, DeKalb, Illinois, USA}
\affiliation{Regis University, Denver, Colorado, USA}
\affiliation{School of Physics and Astronomy, Shanghai Jiao Tong University, Shanghai, China}
\affiliation{Tsung-Dao Lee Institute, Shanghai Jiao Tong University, Shanghai, China}
\affiliation{Department of Physics and Astronomy, Trinity University, San Antonio, Texas, USA}
\affiliation{Institut f\"ur Kern- und Teilchenphysik, Technische Universit\"at Dresden, Dresden, Germany}
\affiliation{Universit\`a del Molise, Campobasso, Italy}
\affiliation{Universit\`a di Udine, Udine, Italy}
\affiliation{Department of Physics and Astronomy, University College London, London, United Kingdom}
\affiliation{University of Illinois at Urbana-Champaign, Urbana, Illinois, USA}
\affiliation{University of Kentucky, Lexington, Kentucky, USA}
\affiliation{University of Liverpool, Liverpool, United Kingdom}
\affiliation{Department of Physics and Astronomy, University of Manchester, Manchester, United Kingdom}
\affiliation{Department of Physics, University of Massachusetts, Amherst, Massachusetts, USA}
\affiliation{University of Michigan, Ann Arbor, Michigan, USA}
\affiliation{University of Mississippi, University, Mississippi, USA}
\affiliation{University of Virginia, Charlottesville, Virginia, USA}
\affiliation{University of Washington, Seattle, Washington, USA}
\affiliation{City University of New York at York College, Jamaica, New York, USA}
\author{D.~P.~Aguillard} \affiliation{University of Michigan, Ann Arbor, Michigan, USA}
\author{T.~Albahri} \affiliation{University of Liverpool, Liverpool, United Kingdom}
\author{D.~Allspach} \affiliation{Fermi National Accelerator Laboratory, Batavia, Illinois, USA}
\author{J.~Annala} \affiliation{Fermi National Accelerator Laboratory, Batavia, Illinois, USA}
\author{K.~Badgley} \affiliation{Fermi National Accelerator Laboratory, Batavia, Illinois, USA}
\author{S.~Bae{\ss}ler} \affiliation{University of Virginia, Charlottesville, Virginia, USA}
\author{I.~Bailey} \altaffiliation[Also at ]{The Cockcroft Institute of Accelerator Science and Technology, Daresbury, United Kingdom.} \affiliation{Lancaster University, Lancaster, United Kingdom}
\author{L.~Bailey} \affiliation{Department of Physics and Astronomy, University College London, London, United Kingdom}
\author{E.~Barlas-Yucel} \altaffiliation[Now at ]{Virginia Tech, Blacksburg, Virginia, USA.} \affiliation{University of Illinois at Urbana-Champaign, Urbana, Illinois, USA}
\author{T.~Barrett} \affiliation{Cornell University, Ithaca, New York, USA}
% ...
\footnotetext[2]{Deceased.}
\renewcommand{\thefootnote}{\arabic{footnote}}
\collaboration{The Muon \gmtwo Collaboration} \noaffiliation
\vskip 0.25cm

\date{\today}

\begin{abstract}
A new measurement of the magnetic anomaly $\amu$ of the positive muon is presented based on data taken from 2020 to 2023 by the Muon \gmtwo Experiment at \ac{FNAL}. This dataset contains over 2.5 times the total statistics of our previous results.  From the ratio of the precession frequencies for muons and protons in our storage ring magnetic field, together with precisely known ratios of fundamental constants, we determine $\amu = \amuRunFourFiveSix$~(\amuprecisionFourFiveSixppb) for the new datasets, and  $\amu = \amuFNAL$~(\amuFNALprecisionppb) when combined with our previous results. The new experimental world average, dominated by the measurements at FNAL, is $a_\mu (\text{exp}) = \amuworld$~(\amuworldprecisionppb). The measurements at \ac{FNAL} have improved the precision on the world average by over a factor of four.
\end{abstract}

\maketitle

\textit{Introduction} --- Precise measurements of magnetic moments of charged leptons serve as precision probes of the \ac{SM} due to their sensitivity to particles and interactions within the \ac{SM} and potentially \ac{BSM}. The Dirac equation~\cite{Dirac1928} predicted $g^{}_\text{e} \equiv 2$ for the g-factor $g^{}_\text{e}$ that relates the electron magnetic moment to its spin.
Schwinger's radiative correction~\cite{Schwinger1948}, inspired by contemporaneous experimental data~\cite{PhysRev.72.1256.2,Kusch1948}, refined this result and introduced the anomaly $a^{}_\text{e}=\alpha/2\pi$. This work laid the foundation for modern relativistic field theory and the development of the \ac{SM}.

The magnetic anomaly $a \equiv (g - 2)/2$~\footnote{The magnetic anomaly $a$ is often also called the anomalous magnetic moment.} arises from radiative corrections from virtual particles and can be calculated precisely within the \ac{SM}. While $a^{}_{\text{e}}$ is measured~\cite{Fan:2022eto} 1000 times more precisely than \amu, the muon's greater mass makes \amu about $4\times10^4$ times more sensitive to much \ac{BSM} physics. Precision measurements of $g^{}_{\mu}$ span decades of advances, beginning with early experiments at Columbia University Nevis Laboratory~\cite{PhysRev.109.973,PhysRev.118.271} and the University of Liverpool~\cite{Cassels_1957}. Direct measurement of \amu started with the CERN-I~\cite{Charpak1965}, CERN-II~\cite{BAILEY1968287} and CERN-III experiments~\cite{Bailey1979}, which the \ac{BNL} E821 experiment further improved~\cite{BNLFinalReport}. The E821 results revealed a statistically significant tension with \ac{SM} predictions at the time. The Muon \gm Experiment at \ac{FNAL} confirmed the E821 result with the 2018 \RunOne data~\cite{Run1PRL}, then refined \amu with over twice the precision  with the \RunTwoThree data~\cite{Run23PRL}.

This paper presents a measurement of \amu from the Muon \gm Experiment using data collected in three runs spanning 2020 to 2023 (designated as \RunFour, \RunFive, and \RunSix). The \RunFourFiveSix positron statistics, over 2.5 times that of our previous measurements~\cite{Run1PRL, Run23PRL}, improve our final \RunOneToSix statistical precision by more than 2.5.  Our final result surpasses our original statistical and systematic goals~\cite{TDR} and establishes a stringent benchmark for future theoretical \ac{BSM} extensions. 

\textit{Experimental principle} --- Our \RunOne and \RunTwoThree publications ~\cite{Run1PRL,Run1PRDomegaa,Run1PRAField,Run1PRABBeamDyn, Run23PRL, Run23PRD} detail the experiment.
Polarized muon beams are injected into a \SI{7.112}{m} radius storage ring with a design storage momentum of ~\SI{3.1}{\GeV/c}.  A superconducting magnet generates a homogeneous vertical \SI{1.45}{T} dipole field~\cite{Danby:2001eh} that provides weak horizontal focusing of the beam and drives the muon spin precession.  Two critical components for beam storage are a fast kicker that redirects muons onto the central orbit~\cite{kickerpaper} and an \ac{ESQ} system for vertical focusing~\cite{e821quadpaper}. At the design momentum, the contributions to the muon spin precession from the electric fields in the \ac{ESQ} cancel.

The experiment determines the ratio of two frequencies, $\Rmu^{'} =\wa/\opprimetildeatTexp$, where \wa is the difference between the spin precession and cyclotron frequencies of the muon, and \opprimetildeatTexp is the \ac{NMR} precession frequency of shielded protons in a spherical water sample (corrected to a reference temperature \Tr), averaged over the muon distribution, which expresses the magnetic field strength. The \wa measurement utilizes 24 PbF$^{}_2$ electromagnetic (EM) calorimeters \cite{Kaspar_2017,KHAW2019162558,ANASTASI201786} that record the energy and time of incident positrons. Parity violation in muon decay and the Lorentz boost of the beam couple to provide an oscillation in the rate of high-energy positrons at a frequency of \wa. A laser system~\cite{Anastasi_2019} continuously monitors the gain of each crystal in the calorimeters. A chain of magnetic field measurements yields \opprimetildeatTexp, where the tilde indicates muon-weighted averaging. The chain begins with a periodic mapping of the magnetic field by a movable mapper with $17$ \ac{NMR} probes \cite{Corrodi_2020}. These probes are calibrated {\it  in-situ} against a water-based cylindrical probe that transfers the absolute calibration of shielded protons in a spherical water sample~\cite{Flay_2021}. Additional \ac{NMR} probes~\cite{swanson2025fixedprobestoragering}, embedded in the experiment's vacuum chambers, track the field while muons are stored between mappings. The mapped and tracked magnetic field is weighted by the  muon distribution, $M$, measured using two straw tracker stations \cite{trackerpaper}.

The narrow aperture through which the beam enters the storage ring produces a mismatch between the phase space of the incoming beam and the storage ring acceptance that leads to \ac{CBO}.
This coherent beam motion introduces a time variation into the positron detection efficiency.
After \RunFour, an additional RF system \cite{Kim_2020}  added a small modulation of the \ac{ESQ} high voltage during the first \SI{6}{\micro\second} after muon injection. The RF system generates dipole fields tuned to the CBO frequency, resonantly damping the CBO by applying forces out of phase.
For analysis purposes, the data is divided into four distinct datasets based on RF configurations: noRF (no RF system), xRF (horizontal RF fields only), and xyRF5/xyRF6 (both horizontal and vertical RF fields in Run-5 and Run-6, respectively).

The measured anomalous spin frequency, \wam, and the muon-weighted magnetic field, $\langle \opprime\times M\rangle$,  must be corrected for several effects via
\begin{eqnarray} \label{eq:R}
   \mathcal{R}_\mu' = \frac{\wam \left(1+C^{}_{e}+C^{}_{p}+C^{}_{pa}+C^{}_{dd}+C^{}_{m l}\right)}{\langle \opprimeatTexp\times M\rangle(1+B^{}_k+B^{}_q )},
\end{eqnarray}
collectively shifting $\Rmu^{'}$ by \SI{572}{ppb} (see Tab.~\ref{tb:systematics}). The corrections to \wam address: the residual contribution to the muon spin precession rate from electric fields $C_e$; the contribution to the muon spin precession from the vertical betatron motion $C_p$; time-dependent changes in the mean phase of the observed muon ensemble caused by (i) detector acceptance $C_{pa}$, and by phase-momentum correlations coupled to (ii) momentum-dependent muon lifetimes $C_{dd}$ and (iii) momentum-dependent muon storage losses $C_{ml}$. Corrections to the muon-weighted magnetic field accommodate the fast transient fields not captured in the \ac{NMR}-based field maps, specifically from eddy currents generated by the fast injection kickers $B_k$ and from vibrations of the \ac{ESQ} plates $B_q$, both synchronous with muon injection. Table~\ref{tb:systematics} summarizes the corrections. 

From $\Rmu^{'}=\wa/\opprimetildeatTexp$, we determine \amu via: 
\begin{linenomath*}
\begin{equation}
\begin{aligned}
a_{\mu} = \frac{\wa}{\opprimetildeatTexp} \frac{\mu'_\text{p}(\Tr)}{\mu^{}_\text{B}} \frac{m_{\mu}}{m_\text{e}},
\label{eq:amueqnewcodata}
\end{aligned}
\end{equation}
\end{linenomath*}
where $\mu'_p(\Tr)/\mu^{}_B$ is the ratio of the shielded proton magnetic moment to the Bohr magneton at $\Tr = \SI{25}{\celsius}$ and $m_{\mu}/m_e$ is the muon-to-electron mass ratio~\footnote{$\mu'_p/\mu^{}_B=\SI{1.5209931551(62)e-3}{}$;\newline $m_{\mu}/m_e=206.768\,2827(46)$, both from~\cite{CODATA:2024}.}.

\begin{table}
\begin{ruledtabular}
\begin{tabular}{lrr}
\multirow{2}{*}{Quantity} & Correction & Uncertainty\\
                          & (ppb)            & (ppb)\\
\hline
\wam\ (statistical) & $\cdots$ & 114\Tstrut\\
\wam\ (systematic)  & $\cdots$ & 30\Bstrut\\
\hline
$C_e$ & 347 & 27 \\
$C_{p}$ & 175 & 9 \\
$C_{pa}$ & -33 & 15 \\
$C_{dd}$ & 26 & 27 \\
$C_{ml}$ & 0 & 2 \\
\hline
$\langle \opprime\times M\rangle$ (mapping, tracking) & $\cdots$  & 34\\
$\langle \opprime\times M\rangle$ (calibration) & $\cdots$ & 34\\
$B_k$ & -37 & 22 \\
$B_q$ & -21 & 20 \\
\hline
\hline
$\mu'_p/\mu_B$  & $\cdots$ & 4\Tstrut\\
$m_\mu/m_e$  & $\cdots$ & 22\\
\hline
Total systematic for \Rmuprime & $\cdots$ & 76\Tstrut\\
Total for \amu & 572 & 139\Bstrut\\
\end{tabular}
\end{ruledtabular}
\caption{Values and uncertainties of the \Rmuprime terms in Eq.~\eqref{eq:R}, and uncertainties due to the external parameters in Eq.~\eqref{eq:amueqnewcodata} for \amu. 
The $\langle \opprime\times M\rangle$ uncertainties are separated into mapping and calibration contributions.
\label{tb:systematics}}
\end{table}

\textit{Anomalous precession frequency }\wam --- Reconstruction of either individual positron candidates or the total calorimeter energy from the digitized EM calorimeter waveforms provides the oscillating time series that get fit to determine \wam.
Four independent analysis groups utilize three different positron reconstruction and pileup correction algorithms while a fifth group performs energy reconstruction~\cite{Run1PRDomegaa,Run23PRD}.  The positron reconstructions have nonlinear energy differences due to the treatment of low energy crystals, differ in the level of unresolved overlapping positrons (pileup) by up to an order of magnitude, and differ in the positron content by several percent.  Consistency of the \wam values from these algorithms speaks to the robustness of the measurement.

Each positron-based analysis constructs its time series using one of three methods~\cite{Run1PRDomegaa,Run23PRD}. The first method bins the positron time directly. The second subdivides the data to construct a ratio that isolates the oscillation from the exponential decay and reduces sensitivity to slow response changes. A new method fits for \wam in slices of the dominant horizontal CBO phase to reduce sensitivity to \ac{CBO}.  Parity violation in muon decay manifests as an energy-dependent amplitude, or asymmetry, $A(E_e)$ in the \wam oscillation. Weighting each positron by $A(E_{e})$ maximizes the statistical power~\cite{Bennett:2007zzb}. The five groups provide seven asymmetry-weighted results for $E_{e}>\SI{1.0}{GeV}$ whose simple average (assuming full correlation) gives \wam. Two energy-based and 11 unit-weighted measurements for $E_{e}>\SI{1.7}{GeV}$  provide 13 crosschecks.

\begin{figure}[ht]
\centering
\includegraphics[width=\columnwidth]{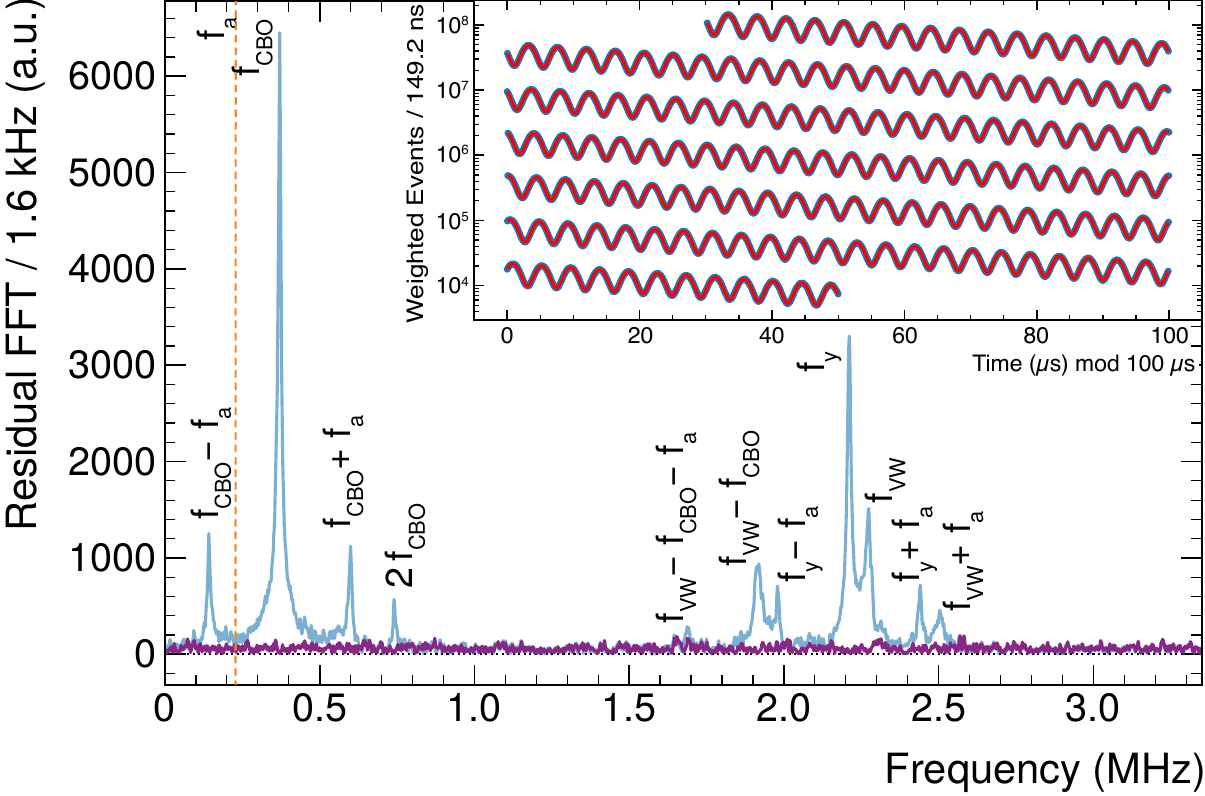}
\caption{Fourier transform of the residuals from the sum of the four fits to the \RunFourFiveSix data excluding (blue) and including (purple) the $\xi(i)$ terms that incorporate the beam oscillation effects.  The peaks correspond to the betatron frequencies (see~\cite{Run1PRDomegaa} for frequency definitions).  The RF-driven CBO damping has decreased the power at $f_{\rm CBO}$ compared to earlier data.  The dashed line (orange) indicates the anomalous precession frequency $f_a$.
Inset: the asymmetry-weighted $e^{+}$ time spectrum for the summed data (blue) and fit functions (red).}
\label{fig:ffts}
\end{figure}

Each group fits their time series using one of two models. The first, used previously, takes the form
\begin{linenomath*}
\begin{multline}
N(t)=N_0 \Lambda(t) \eta_{N}(t) e^{-t / \gamma\tau_\mu} \\
\times\left\{1+A \eta_{A}(t)\cos\left[\wam t - \varphi_0 + \eta_{\phi}(t)\right]\right\},
	\label{eq:wiggle_func1}
\end{multline}
\end{linenomath*}
where $N_{0}$ is the normalization, $\gamma\tau_\mu$ is the average boosted muon lifetime ($\approx\SI{64.4}{\micro\second}$), $\Lambda(t)$ accounts for beam loss, $A$ is the average rate asymmetry, and $-\varphi_0$ is the average phase extrapolated to $t=0$.  The factors $\eta_N$, $\eta_A$,  and
$\eta_\phi$ are well-motivated corrections that accommodate the rate variations from the coupling of calorimeter acceptance with beam oscillations, whose frequencies can be seen in Figure~\ref{fig:ffts}. A second, complementary model replaces those factors with a sum $\sum_{\omega_{i}} \xi_{i}(t)$ over those frequencies that modulate the $\{1+A\cos(\wam t - \varphi_0)\}$ precession term.
 The fits begin near \SI{30}{\micro\second} to allow the \ac{ESQ} fields, and thus the muon beam,  to stabilize.

Each analysis group developed its analysis using an unknown, fixed, pseudorandom offset in the precession frequency.
These blindings add to the hardware blinding of the digitization frequency, which was set and monitored by FNAL physicists outside the collaboration.  Shortly before the hardware unblinding, the groups shifted to a common unknown blind and two of the seven analyses included in the final average addressed minor issues in the $\eta_{\phi}(t)$ treatment exposed by the comparisons, with changes on the scale of a few \ac{ppb}.  Removal of the hardware blinding occurred after all aspects of the \amu determination were complete and frozen.

All fits model the data well, and meet the requirement for inclusion in the final result:  a good $\chi^{2}$ and a Fourier transform of the fit residuals free of artifacts, as illustrated for the combined data in Fig.~\ref{fig:ffts}. The fit for the highest statistics dataset has a $\chi^2$ of 4007 for 4097 degrees of freedom, and a probability of $\chi^2$ of 84\%.  Fits in positron energy bins, individual calorimeter stations, and as a function of fit start and end time show only statistical scatter in \wam.  The twenty correlated \wam values agreed within the allowed statistical variations assessed using 200 bootstrap samples.
The \wam statistical uncertainty of \SI{114}{ppb} dominates the \RunFourFiveSix  \amu uncertainty. 

Uncertainties in the envelopes of the transverse beam oscillations and in a new correction for an intensity-dependent calorimeter gain sag dominate the systematic uncertainties, each contributing at the level of 25-40~ppb in the four fit datasets.
The gain sag largely explains a residual slow term observed in the data in the previous publications~\cite{Run1PRDomegaa,Run23PRD} (in addition to a small issue in one of the reconstructions that has been corrected here), and has been well-measured with dedicated laser studies using a full calorimeter station.  While the magnitude of the sag is below our design specification, it oscillates at \wam with a phase shift relative to the beam intensity, creating a greater \wam sensitivity than previously estimated.  

\textit{Beam-dynamics corrections ${C_i}$} ---
The leading correction $C_e$ to $\wam$ derives from  the measured muon momentum distribution, which has about a \SI{0.1}{\percent} relative width.
The debunching of the muon beam that results from the spread of cyclotron frequencies can be observed in calorimeter data at early times in a muon fill. That data determines the momentum distribution of the stored muons and the correlation between momentum and relative time in the bunch~\cite{Run1PRABBeamDyn,Run23PRD}. The reconstructed radial muon distribution from the trackers also determines that information. Both methods were cross-checked with a minimally-intrusive, insertable scintillating fiber detector in dedicated studies, a novelty of this dataset. The staged introduction of the \ac{ESQ} RF reduced the size of this correction from \SI{387}{ppb} to \SIrange{318}{382}{ppb}. The improved understanding and enhanced robustness of these methods reduced the uncertainty from \SI{32}{ppb} in \RunTwoThree to \SI{27}{ppb} in \RunFourFiveSix.

The tracker data provide the distribution of the stored muons' vertical betatron amplitudes, which lead to the correction $C_p$. The analysis method remains unchanged from the \RunOne~\cite{Run1PRABBeamDyn} and \RunTwoThree~\cite{Run23PRD} analyses.

As the muon beam evolves, changes in the distribution of muons whose daughter positron has been detected can lead to a time-dependent change in the ensemble-averaged spin precession phase, which causes bias in \wam. The calorimeter acceptance and relative phase versus transverse decay position are evaluated through simulation. Combining these with the transverse beam distribution measured in the trackers provides the $C_{pa}$ correction. The method, resulting values, and uncertainty remain consistent with the \RunTwoThree evaluation.

Correlations between the muon phase and its momentum, together with the momentum dependence of its lifetime, lead to the differential decay correction $C_{dd}$. Phase-momentum correlations in the stored muon ensemble originate from the upstream beamline and the injection process. These effects are evaluated in simulation and supported by dedicated measurements in which the momentum acceptance of the storage ring is changed.  Another contribution arises from correlations of injection time and momentum -- early muons spend additional time in the storage ring field compared to late ones,  so build up a relative precession phase shift. The temporal shape of the kicker pulse and the relative timing between kicker pulse and muon injection dominate those correlations. This contribution is evaluated from the momentum-injection time distributions from calorimeter data that were used to calculate $C_e$. These kicker-related contributions vary over a range of $\pm\SI{50}{ppb}$ for different time periods and for the muon bunch position in the $16$-bunch cycle. Cancellations in averaging  lead to \SIrange{4}{8}{ppb} corrections to the main datasets with uncertainties of \SIrange{17}{19}{ppb}.

Lastly, time-dependent phase changes can be caused by muon loss. Loss rates have been significantly reduced compared to earlier runs by optimizing storage conditions. The introduction of RF decreased these losses further, leaving a negligibly small correction, $C_{ml}$.

\begin{figure}[ht]
\centering
\includegraphics[width=1.0\columnwidth]{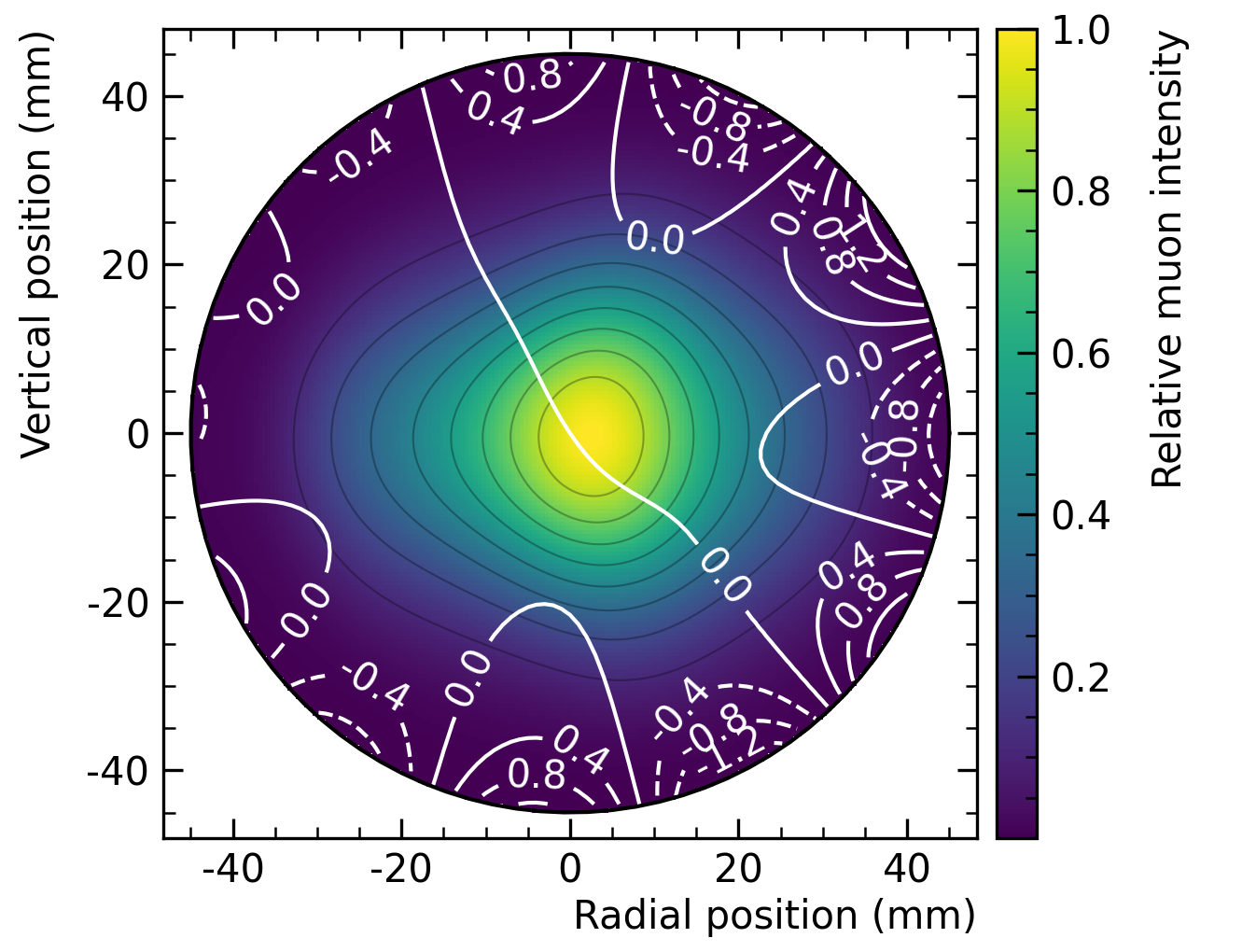}
\caption{Azimuthally averaged magnetic field contours in units of \SI{}{ppm} overlaid in white on the time- and azimuthally averaged muon distribution for the xyRF5 dataset.} \label{fig:field}
\end{figure}

\textit{Muon-weighted magnetic field \weightedfield} ---
The determination of the muon-weighted magnetic field largely follows the procedure developed earlier~\cite{Run1PRAField,Run23PRD}.
The magnetic field showed azimuthal variations with an RMS below \SI{20}{ppm}. The relative field change over the muon storage radius of \SI{45}{\milli\meter} caused by azimuthally-averaged transverse linear and higher-order gradients is below \SI{100}{ppb} and \SI{1}{ppm}, respectively.
Figure~\ref{fig:field} superimposes the azimuthally averaged magnetic field contour lines from the xyRF5 dataset on the time- and azimuthally averaged muon distribution.

The \RunFourFiveSix dataset spans $194$ magnetic field maps, compared with $16$ in \RunOne and $69$ for \RunTwoThree.
All but four of the azimuthally averaged magnetic maps field stayed within a range of $\pm\SI{1.3}{ppm}$. The remainder, still included in this work, exhibited variations up to \SI{10}{ppm}.

The uncertainty in the field maps for \RunFourFiveSix reduced to \SIrange{29}{31}{ppb} compared to \SIrange{37}{39}{ppb} in \RunTwoThree, primarily due to additional measurements of the magnetic footprint of the field mapper's parking mechanism. Its uncertainty contribution dropped from \SI{22}{ppb} to \SI{2}{ppb}, but these map uncertainty gains are partially offset by the larger temperature differences between the \ac{NMR} probes during magnetic field mapping and during their {\it in-situ} calibration, increasing the uncertainty from the \ac{NMR} probe temperature dependence from \SIrange{9}{15}{ppb} to \SIrange{13}{16}{ppb}.

The increased number of tracked magnetic field maps compared to our previous publications allowed for more detailed studies of systematic effects in that tracking, particularly of local drifts at magnet pole edges as a function of time after a magnet ramp-up. The remaining systematic uncertainty is evaluated using a Brownian bridge model and a new 
time-independent constant model, which yielded similar results. The associated total tracking uncertainty reduced to \SIrange{8}{14}{ppb} from \SIrange{17}{18}{ppb} due to the larger number of magnetic field maps.
The muon weighting, which adjusts the field maps for the field experienced on average by the stored muons, follows our previous approach, yielding uncertainties of \SIrange{6}{7}{ppb} comparable to those of \RunThreeB which had similar muon distributions.

Two analysis teams with separate codebases performed the magnetic field map extraction, tracking, and weighting processes while software blinded. Their results were consistent. The uncertainties of magnetic field mapping, tracking and muon weighting combine for a total of \SIrange{33}{35}{ppb} by dataset and \SI{34}{ppb} in the final average. 

{\em In-situ} calibrations of the field mapper NMR probes were performed before and after each of the three \RunFourFiveSix running periods~\footnote{The calibration after \RunSix is only used as a cross-check and doesn't enter the average.}. No time dependence was observed over the six years of data-taking, and the per-probe calibrations are averaged, contributing an uncertainty of \SI{16}{ppb}. Since the same calibration probe was used in \RunTwoThree, the absolute calibration with respect to shielded protons in a spherical water sample remains unchanged. Additional work enhanced our understanding of the material effects of the calibration probe but had no direct impact on the associated \SI{8}{ppb} uncertainty.

The absolute calibration was cross-checked against $^{3}$He-based magnetometers~\cite{PhysRevLett.124.223001}, which showed a \SI{1.7}{\text{standard deviation}} difference. Comparisons to water-based continuous-wave \ac{NMR} probes developed by the J-PARC Muon \gm/EDM collaboration~\cite{8626508} showed inconsistencies in early comparisons but good agreement in later iterations. These new crosschecks lead to an additional \SI{25}{ppb} uncertainty, resulting in a total calibration uncertainty of \SI{34}{ ppb}.

The total systematic uncertainty of the muon-weighted magnetic field, including calibration, increased by \SI{10}{ppb} to a total of \SI{48}{ppb} with respect to \RunTwoThree, primarily driven by the uncertainties on the absolute calibration.

\textit{Magnetic field transients $B_i$} ---
The time-dependent residual magnetic fields from eddy currents induced by the fast kicker magnets were measured {\it in-situ} with two independent magnetometers, both based on Faraday rotation in terbium gallium garnet crystals. One system uses optical fibers to guide the laser light into the kicker region. The second new system uses an open laser beam, which enters the storage volume through a window, and mirrors to guide the light.
Two of the three kickers were measured at several transverse positions. The measurements from the two apparatus agree within uncertainties.
Measurements at a large radius (\SI{+18}{mm}) revealed transverse variations of about a factor of two, which were investigated with additional measurements on a mock-up. While the measurements close to the beam center are consistent with earlier determinations, the new measurements at a larger radial offset, together with improved understanding of the transverse modeling of the resulting field perturbation, lead to a larger correction term $B_k$ than reported in \RunTwoThree. Driven by the observed stronger transverse variation of the transient and uncertainties in the transverse modeling, the total uncertainty increased to \SI{22}{ppb} compared to \RunTwoThree.

The correction presented in reference~\cite{Run23PRD} for magnetic field transients from vibrations in the \ac{ESQ} system, $B_{q}$, remains valid. The transverse distribution of this effect was studied in more detail, and the assigned uncertainties were corroborated.

\textit{Calculation of $a_\mu$} ---
The values of \wa and \opprimetilde listed in Tab.~\ref{tb:Rmu} for the four fit datasets, along with the corrections from Eq.~\eqref{eq:R}, form \Rmuprime.  The  \Rmuprime measurements are statistically uncorrelated, while nearly all systematic uncertainties are fully correlated. The table also lists the combined \RunFourFiveSix result and our earlier results, corrected as noted below.
The four fit datasets show good consistency with a $\chi^2=0.96$ for 3 degrees of freedom, which has a probability of \SI{80}{\percent}. No statistically significant correlations with magnet current, magnetic field, field gradients, or time of day were observed.

We report \Rmuprime at $\Tr=\SI{25}{\celsius}$.  This change from the reference temperature used in previous publications aligns with both the CODATA standard and our actual measurement conditions. Our previous \Rmuprime values were adjusted by \SI{-101}{ppb} to reflect this change in reference temperature and external constants. The \amu values do not change as a result of the \Tr shift, though the external CODATA constants have been updated.

The superior statistical power of this larger dataset, along with additional dedicated measurements, enabled further cross-checks of the \RunTwoThree results. Three corrections with corresponding uncertainty adjustments were identified and applied when combined with the latest dataset: the sensitivity of \wa to small, slow gain shifts noted earlier, improved understanding of spatial dependencies in the transient magnetic fields from kicker system eddy currents, and a sign error correction in one component of the $C^{}_{dd}$ correction. These corrections, determined independently, happened to have the same sign and combine to shift the \RunOne and \RunTwoThree results by \SI{50}{ppb} and \SI{98}{ppb}, respectively, and result in a total systematic uncertainty of \SI{78}{ppb} for the adjusted \RunTwoThree result.  The corrections were finalized before unblinding the \RunFourFiveSix results.
The latest result agrees well with the previous measurements.

The combined FNAL average, \RunOneToSix, with a total uncertainty of \amuFNALprecisionppb, assumes fully correlated systematic uncertainties between the results.

\begin{figure}[ht]
\centering
\includegraphics[width=\columnwidth]{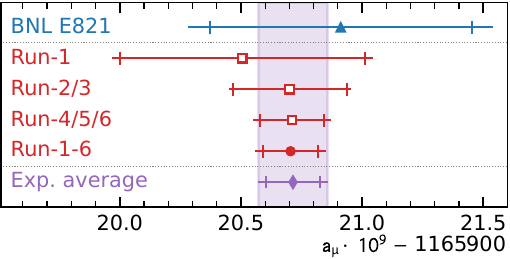}
\caption{Experimental values of \amu from BNL E821~\cite{BNLFinalReport} (blue triangle), our \RunOne~\cite{Run1PRL}, \RunTwoThree~\cite{Run23PRL} and \RunFourFiveSix  (red squares), those three results combined (red circle), and the new experimental world average (purple diamond). The inner tick marks indicate the statistical contribution to the total uncertainties. Corrections to earlier results have been applied.} \label{fig:results}
\end{figure}

Following Eq.~\ref{eq:amueqnewcodata}, we determine the muon anomaly 
\begin{equation*}
\begin{split}
a_\mu({\text{Run-4/5/6}}) &= \amuRunFourFiveSix  ~~~ (\amuprecisionFourFiveSixppb) \\
a_\mu({\text{Run-1-6}}) &= \amuFNAL  ~~~ (\amuFNALprecisionppb),
\end{split}
\end{equation*}for the full dataset, with the statistical, systematic, and external parameter uncertainties combined in quadrature.
The combined experimental (exp) average, from BNL E821~\footnote{The adjusted  $\Rmuprime(\text{BNL})=0.00370730154(172)_{\text{stat.}}(102)_{\text{syst}}$ contributes \SI{5.1}{\percent} to the new experimental average.} and Run-1-6, becomes
\begin{equation*}
\amu(\text{exp}) = \amuworld   ~~~(\amuworldprecisionppb).
\end{equation*}Figure~\ref{fig:results} shows the corresponding values.

The Muon \gm Theory Initiative has released an updated \ac{SM} value of $a_\mu$ in their 2025 White Paper (WP2025)~\cite{WP25}, based on results from~\cite{\SMref}, which agrees with the measured average. The value shifts considerably compared to their 2020 White Paper (WP2020)~\cite{TI}, which is almost entirely due to the exclusive use of new, published leading-order hadronic vacuum polarization estimates based on lattice-QCD calculations. 
The previous value in their WP2020 used experimental $e^+e^-\rightarrow$ hadrons cross-section measurements from multiple experiments to evaluate this contribution based on a dispersion integral and showed a discrepancy with the experimental value. However, a recent cross-section measurement \cite{PhysRevD.109.112002,PhysRevLett.132.231903} has increased the tension among the experimental inputs, thus a prediction based on the dispersion integral was not included in their WP2025. Efforts are continuing towards an evaluation of this leading-order hadronic contribution using both lattice-QCD and dispersion integral calculations.

\begin{table}
\begin{ruledtabular}
\begin{tabular}{lrrr@{\!\!\!\!}l}
  & \multicolumn{1}{l}{$[\omega_a/2\pi\,-$} & \multicolumn{1}{l}{[$\opprimetilde/2\pi\,-$} & \multicolumn{2}{l}{$[\Rmuprime \times 10^{11}$\,$-$}\\
 & \multicolumn{1}{r}{$\SI{229077}{}]$\,(Hz)} & $\SI{61790900}{}]$\,(Hz) & \multicolumn{2}{r}{$\SI{370730000}{}]$} \\
\hline
Run-1 & $\cdots$ & $\cdots$                             & 25&(161)(59)\\
Run-2/3 & $\cdots$ & $\cdots$                           & 87&(75)(29)\\
\hline
noRF & 0.504(42)(12) & 20.0(0.5)(3.5)  &  43&(68)(29)\\
xRF & 0.626(55)(11) & 38.9(0.7)(3.4)   & 126&(90)(28)\\
xyRF5 & 0.500(56)(12) & 10.9(0.7)(3.4) &  90&(91)(28)\\
xyRF6 & 0.509(64)(11) & 03.6(0.9)(3.5) & 148&(103)(28)\\
\hline
\multicolumn{2}{l}{Run-4/5/6 \hfill $\cdots$} & $\cdots$        & 90&(42)(28)\\
\hline
Run-1-6 & $\cdots$ &                               $\cdots$ & 88&(36)(29)\\
\end{tabular}
\end{ruledtabular}
\caption{
Measurements of $\wa$, $\opprimetilde$, and their ratios $\Rmuprime$, with $\wa/2\pi$ and $\opprimetilde/2\pi$ values shown as offsets from +\SI{229077}{Hz} and +\SI{61790900}{Hz}, respectively, and $\Rmuprime \times 10^{11}$ values as offsets from \SI{370730000}{}. The \RunOne and \RunTwoThree values have been updated from~\cite{Run23PRL} as described in the text. The uncertainties are shown in the form $()_{\rm{stat.}}()_{\rm{syst.}}$.
}
\label{tb:Rmu}
\end{table}

In summary, we report the measurement of the muon magnetic anomaly to a precision of \amuFNALprecisionppb using our full  six years of data. With over a four-fold improvement in precision over the BNL E821 measurement~\cite{BNLFinalReport}, this result represents the most precise determination of the muon magnetic anomaly and provides a powerful benchmark for extensions of the \ac{SM}.

We thank the Fermilab management and staff for their strong support of this experiment, as well as our university and national laboratory engineers, technicians, and workshops for their tremendous support.
Greg Bock, Joe Lykken, and Rick Ford set the blinding clock and diligently monitored its stability. We also thank members of the J-PARC Muon \gm/EDM experiment for the cross-calibration efforts.

The Muon \gmtwo Experiment was performed at the Fermi National
Accelerator Laboratory, a U.S. Department of Energy, Office of
Science, HEP User Facility. Fermilab is managed by Fermi Forward Discovery Group, LLC, acting under Contract No. 89243024CSC000002.
Additional support for the experiment was provided by the U.S. DOE Office of Science under the offices of HEP, NP, ASCR, and the U.S.-Japan Science and Technology Cooperation Program in HEP, the U.S. National Science Foundation, the Istituto Nazionale di Fisica Nucleare (Italy), the Science
and Technology Facilities Council (UK), the Royal Society (UK),
the National Natural Science Foundation of China
(Grant No. 12475108, 12305217, 12075151), MSIP, NRF, and IBS-R017-D1 (Republic of Korea),
the Deutsche Forschungsgemeinschaft (DFG, German Research Foundation, Germany) through the Cluster of
Excellence PRISMA+ (EXC 2118/1, Project ID 39083149),
the European Union Horizon 2020 research and innovation programme under
the Marie Sk\l{}odowska-Curie Grant Agreements No. 101006726 and No. 734303,
the European Union STRONG 2020 project under Grant Agreement No. 824093,
and the Leverhulme Trust, LIP-2021-014.

\bibliography{PRL-Run456, WP25_essential_inputs}
\begin{acronym}
  \acro{2D}{two-dimensional}
  \acro{3D}{three-dimensional}

\acro{ADC}{analog-to-digital converter}  
  \acro{ANL}{Argonne National Laboratory}

\acro{BNL}{Brookhaven National Laboratory}
  \acro{BSM}{Beyond the Standard Model}
  
\acro{CBO}{coherent betatron oscillation}
  \acro{COD}{closed orbit distortion}
  \acro{CPU}{central processing unit}
  \acro{ctag}{calorimeter tag}
  \acro{CERN}{European Organization for Nuclear Research}

\acro{DAC}{digital-to-analog converter}  
  \acro{DAQ}{data acquisition}  
  \acro{DQC}{data quality cuts}  
  \acro{DQM}{data quality monitoring}  

\acro{ESQ}{electrostatic quadrupole}

\acro{FFT}{fast fourier transform}
  \acro{FID}{free induction decay}
  \acro{FNAL}{Fermi National Accelerator Laboratory}
  \acro{FPGA}{field programmable gate arrays}

\acro{GPS}{global positioning system}
  \acro{GPU}{graphic processing unit}
  \acro{GUI}{graphic user interface}

\acro{HV}{high voltage}
\acro{IBMS}{Injected Beam Monitoring System}
  \acro{IRIG-B}{inter-range instrumentation group code B}
  \acro{IC}{integrated circuit}

\acro{LVDS}{low-voltage differential signaling}
  \acro{LED}{light emmitting diode}

\acro{MIDAS}{maximum integrated data acquisition system}
 \acro{MRI}{magnetic resonance imaging}

\acro{NMR}{nuclear magnetic resonance}

\acro{ODB}{online database}

\acro{ppb}{parts per billion}
  \acro{ppm}{parts per million}
  \acro{ppt}{parts per trillion}
  \acro{PEEK}{polyether ether ketone}
  \acro{PLL}{phase-locked loop}
  \acro{POT}{potentiometer}
  \acro{QCD}{quantum chromodynamics}
  \acro{QED}{quantum electrodynamics}
  \acro{PID}{proportional–integral–derivative}
  \acro{PBSC}{polarizing beam splitter cube}

\acro{RF}{radio frequency}
  \acro{RMS}{root mean square}

\acro{SCC}{surface correction coils}
  \acro{SPI}{serial peripheral interface}
  \acro{SM}{Standard Model}

\acro{TDR}{technical design report}
  \acro{TI}{Texas Instrument}
  \acro{TTL}{transistor–transistor logic}
  \acro{TGG}{terbium gallium garnet}

\acro{UTC}{universal time coordinated}

\acro{VME}{Versa Module European}
  \acro{VTM}{virtual trolley measurement}

\end{acronym}
 \end{document}


\title{Measurement of the Positive Muon Anomalous Magnetic Moment to 127 ppb
  \\
  Supplemental Material
}
\maketitle
We present additional information for the final measurement of the muon magnetic anomaly, \amu, by the Muon \gm Experiment at \acl{FNAL}. Five sections provide background on the run history, a new \ac{RF} system, the fit functions for the extraction of \oa, a residual gain correction, and our improved understanding of the differential decay correction. A forthcoming publication will further detail the analysis and improvements over our earlier publications.

\textit{Run history} --- The Muon \gm Experiment collected data over six physics runs, summarized in Table~\ref{tab:conditions}, from 2018 to 2023, with improved operational stability achieved in the later periods. 
\RunOne (2018), the first physics run, had variable operational parameters, including kicker voltages ranging from \SIrange{125}{137}{kV} and \ac{ESQ} performance variations due to damaged high-voltage resistors, which were identified and replaced after the run.
During \RunTwoThree (2019-2020) the experiment's temperature stability improved, kicker upgrades enabled better beam centering, and the \ac{ESQ} voltage optimization minimized resonances and reduced muon losses.  The final three runs, the focus of this publication, \RunFourFiveSix, operated under largely consistent conditions, with its primary distinction being the staged introduction of \ac{RF} modulation to the \ac{ESQ} system to reduce \ac{CBO} amplitudes.

\begin{table*}[tb]
    \centering
    \begin{ruledtabular}
    \begin{tabular}{lccccccccc}
         Dataset & ESQ (kV) & RF mode & Kicker (kV) & Field index & Positrons ($10^9$) & Mag. Field Maps \\
         \hline
         \RunOne       & 18.3 & -- & 125 to 137 &  0.107 to 0.120 & 15.4 & 16\\
         \RunTwoThree  & 18.2 & -- & 142 and 161 & 0.107 to 0.108 & 70.9 & 69\\
         \hline
         noRF         & 18.2 & -- & 161 & 0.108 & 86.0 & 71\\
         xRF          & 18.2 & x & 161 & 0.108 & 49.3 & 40\\
         xyRF5        & 18.2 & xy & 161 & 0.108 & 47.8 & 37\\
         xyRF6        & 18.2 & xy & 161 & 0.108 & 39.1 & 46     \\
    \end{tabular}
    \caption{Key parameters for the Muon g-2 dataset periods. The RF mode indicates which \ac{ESQ} RF components were used: no RF (--), horizontal only (x), or both horizontal and vertical (xy), the field index $n$, the number of analyzed positrons within the energy range \SIrange{1}{3.1}{GeV/c} detected more than \SI{30}{\micro\second} after the muon injection, and the number of magnetic field maps.}
    \label{tab:conditions}
    \end{ruledtabular}
\end{table*}

\textit{RF system} ---The \ac{ESQ} \ac{RF} system applied a  $\sim$1\,kV RF voltage modulation tuned to the \ac{CBO} frequency during the first 6\,$\mathrm{\upmu s}$ after beam injection, which resonantly damped the coherent beam motion by more than a factor of five, significantly reducing its impact on \wam. Omission of the CBO-related terms from the \wa fits shifts \wa by 0.8\,ppm in data without the RF but only by 0.1\,ppm with the RF. While the dominant \ac{CBO} systematic uncertainty, stemming from limited knowledge of its decoherence behavior, remained similar between datasets, the consistent \wam values obtained with and without \ac{RF}  bolsters confidence in the robustness of the \ac{CBO} treatment\lkgr{, regardless of amplitude or direct influence on \simon{\wa}{$\omega_{a}^{m}$}}{}. 

\lkgr{Beyond damping the \ac{CBO}, t}{} The \ac{RF} field also helped center the stored beam by rebalancing the phase-space distribution of high- and low-momentum muons, thereby reducing the electric field correction $C_e$ as follows.  Underkicking by the kicker causes the stored muons to oscillate with large \ac{CBO} amplitude. In the absence of \ac{RF}, high-momentum muons, which are initially closer to the ideal orbit, receive a kick closer to optimal and thus remain nearer the ideal orbit, while low-momentum muons get pushed further from equilibrium. As a result, more high-momentum muons survive post-kick, shifting the beam centroid. The RF acts to bring low-momentum muons closer to the ideal orbit while pushing high-momentum muons away, making their motion out of phase but matched in amplitude and population. This centering of the beam reduces both $C_e$ and muon loss during storage, and thus improves the overall quality and stability of the stored beam. 
\textit{Fit function details} ---The transverse oscillations of the muon beam and the spread of muon momenta introduce  complexity into the precession data, which the terms  $\eta_N(t)$, $\eta_{A}(t)$ and $\eta_{\phi}(t)$ capture in the fit model
\begin{linenomath*}
\begin{multline}
N(t)=N_0 \Lambda(t) \eta_{N}(t) e^{-t / \gamma\tau_\mu} \\
\times\left\{1+A \eta_{A}(t)\cos\left[\wam t - \varphi_0 + \eta_{\phi}(t)\right]\right\}.
	\label{eq:wiggle_func1}
\end{multline}
\end{linenomath*}
The oscillations couple to the variation of positron acceptance with muon decay position.  For example, the normalization modulation takes the form
\begin{equation}
    \eta_N(t) = X(t) + Y(t) + XY(t),
\end{equation}
where $X(t)=\sum A_i(t)\cos(\omega_i t + \phi_i(t))$
arises from the coupling of the horizontal beam oscillations with 
horizontal acceptance. A similar form for $Y(t)$ arises from the vertical oscillation and acceptance, while $XY(t)$ mixes horizontal and vertical frequencies because of the variation of the vertical acceptance function with horizontal position.  Nonlinearities in the \ac{ESQ} fields cause a slow drift in the muon ensemble-averaged betatron frequencies over the course of a muon fill which the time dependence in $\phi_i(t)$ captures.  The relative amplitudes of the betatron frequencies are well understood.  The dominantly linear horizontal and quadratic vertical acceptances cause the horizontal betatron frequency, aliased to $f_{\rm CBO}$ and the second harmonic of the vertical frequency, aliased to $f_{\rm VW}$ to dominate. The acceptance modulation also results in modulation of the positron spectrum, which in turn modulates both the muon ensemble-averaged asymmetry $A$ and precession phase $\phi$.  The correction terms $\eta_{A}(t)$ and $\eta_{\phi}(t)$ have forms similar to the $X(t)$ terms.

The five different analyses take a broad range of approaches for incorporating the time dependence of the envelopes $A_i(t)$ or $\alpha_i(t)$ and $\beta_i(t)$, which eventually decay away as the beam oscillations decohere.  The approaches range from purely analytical forms with parameters floating in the fit to data-driven forms where the amplitudes are measured in time bins and then smoothed. 
\textit{Residual gain correction} ---The initial flash of beam particles at injection causes a gain sag in the calorimeters as capacitors are depleted and bias voltages recover. 
The laser system \cite{Anastasi_2019} measures this effect \textit{in-situ} and provides gain corrections that are exponential with $\mathcal{O}(\SI{6}{\mu s})$ time constants governed by the capacitance of the detector electronics~\cite{Run1PRDomegaa,Run23PRD}.
Short term  --- $\mathcal{O}(\SI{10}{ns})$ --- effects from the recovery of individual silicon photomultiplier pixels from single positron hits are also corrected.

Each shower also causes a small gain sag with the $\mathcal{O}(\SI{6}{\mu s})$ time constant.
This effect and its scaling with the number of photoelectrons was confirmed with dedicated measurements of a modified calorimeter station after the completion of \RunSix.
These small gain perturbations build up iteratively over a muon fill and distort the exponential flash recovery. 
The distortion due to this effect was considered during design of the calorimeter electronics \cite{Kaspar_2017} and is visible in the residuals of laser monitor data after the correction for the flash.
The relative amplitude of this effect is below our design goal of $ < 10^{-4}$ at \SI{30}{\micro\second}, so no correction was calculated for previous publications.
    
Because the amplitude of the gain sag is proportional to the deposited energy and the recovery time constant is about the $\omega_a$ period, the oscillation of the average positron energy in the detectors imprints a phase-shifted $\omega_a$ oscillation onto the gain correction. 
This phase shift leads to a bias in \wam much larger than an exponential or in-phase oscillating gain sag if not corrected. With the \RunFourFiveSix beam rates, this gain sag caused a shift of \SI{40}{ppb} in \wam  with an amplitude below the design sensitivity of our laser monitoring system. 
 
\textit{Differential decay} --- The differential decay correction 
\begin{equation} \label{eq:Cdd}
    C_{dd} = \left(\frac{1}{\omega_{a}}\right)\left(\frac{d\varphi_0}{dp}\right)\left(\frac{dp}{dt}\right)_{dd}
\end{equation}
to $\omega_a^m$ arises from correlations between the initial phase of the spin precession after injection and the momentum of stored muons. The coupling of this correlation and the evolving momentum distribution due to the momentum-dependent muon lifetime leads to a temporal evolution of the ensemble-averaged initial phase, $\varphi_0$, which in turn shifts the measured spin precession frequency. 

We directly calculate the time evolution of the momentum spread from its width $\sigma_p$, i.e.,
 \begin{equation}
 \left(\frac{dp}{dt}\right)_{dd} \approx \frac{1}{\bar{p}\gamma\tau_{\mu}}\sigma^2_{p},
  \end{equation}
with $\bar{p}$ the mean momentum.  The momentum-phase correlation $d\varphi_0/dp$ can be decomposed into two parts: the \textit{injection} and \textit{longitudinal} components.

The \textit{injection} component emerges from initial phase correlations with muon momentum and transverse phase space coordinates that originate from the muon production, transport to the experiment, and injection into the storage ring. We extract this component from high-fidelity simulations that reproduce the effects of pion decay and muon momentum on the spin phase advance, as well as the correlations at injection time between the radial phase space coordinates and momentum for muons that store. Since all these aspects define the injection component, we cover a wide range of configurations that lead to a contribution $C_{dd}$(Injection) of \SIrange{19}{20}{} ppb with a conservative uncertainty estimate of \SI{20}{ppb}.

The \textit{longitudinal} component arises from the dependence of momentum acceptance on the longitudinal coordinate within the stored muon bunch. The temporal shape of the kicker pulse determines this effect. This correspondence between momentum and muon time-of-flight accentuates the muon spin-precession gradient from the head to the tail of the stored beam, which together yield $C_{dd}$(Longitudinal) of \SIrange{4}{8}{} ppb with uncertainties of \SIrange{17}{19}{} ppb.  The total correction is $C_{dd}=26\pm27\,\text{ppb}$. 
\bibliography{PRL-Run456}
\begin{acronym}
  \acro{2D}{two-dimensional}
  \acro{3D}{three-dimensional}

\acro{ADC}{analog-to-digital converter}  
  \acro{ANL}{Argonne National Laboratory}

\acro{BNL}{Brookhaven National Laboratory}
  \acro{BSM}{Beyond the Standard Model}
  
\acro{CBO}{coherent betatron oscillation}
  \acro{COD}{closed orbit distortion}
  \acro{CPU}{central processing unit}
  \acro{ctag}{calorimeter tag}
  \acro{CERN}{European Organization for Nuclear Research}

\acro{DAC}{digital-to-analog converter}  
  \acro{DAQ}{data acquisition}  
  \acro{DQC}{data quality cuts}  
  \acro{DQM}{data quality monitoring}  

\acro{ESQ}{electrostatic quadrupole}

\acro{FFT}{fast fourier transform}
  \acro{FID}{free induction decay}
  \acro{FNAL}{Fermi National Accelerator Laboratory}
  \acro{FPGA}{field programmable gate arrays}

\acro{GPS}{global positioning system}
  \acro{GPU}{graphic processing unit}
  \acro{GUI}{graphic user interface}

\acro{HV}{high voltage}
\acro{IBMS}{Injected Beam Monitoring System}
  \acro{IRIG-B}{inter-range instrumentation group code B}
  \acro{IC}{integrated circuit}

\acro{LVDS}{low-voltage differential signaling}
  \acro{LED}{light emmitting diode}

\acro{MIDAS}{maximum integrated data acquisition system}
 \acro{MRI}{magnetic resonance imaging}

\acro{NMR}{nuclear magnetic resonance}

\acro{ODB}{online database}

\acro{ppb}{parts per billion}
  \acro{ppm}{parts per million}
  \acro{ppt}{parts per trillion}
  \acro{PEEK}{polyether ether ketone}
  \acro{PLL}{phase-locked loop}
  \acro{POT}{potentiometer}
  \acro{QCD}{quantum chromodynamics}
  \acro{QED}{quantum electrodynamics}
  \acro{PID}{proportional–integral–derivative}
  \acro{PBSC}{polarizing beam splitter cube}

\acro{RF}{radio frequency}
  \acro{RMS}{root mean square}

\acro{SCC}{surface correction coils}
  \acro{SPI}{serial peripheral interface}
  \acro{SM}{Standard Model}

\acro{TDR}{technical design report}
  \acro{TI}{Texas Instrument}
  \acro{TTL}{transistor–transistor logic}
  \acro{TGG}{terbium gallium garnet}

\acro{UTC}{universal time coordinated}

\acro{VME}{Versa Module European}
  \acro{VTM}{virtual trolley measurement}

\end{acronym}